\documentclass[twocolumn,showpacs,preprintnumbers,amsmath,amssymb]{revtex4}
\usepackage{graphicx}
\usepackage{dcolumn}
\usepackage{bm}
\usepackage{tensor}
\begin{document}
\title{Cold matter trapping via slowly rotating helical potential.}
\author{A.Yu.Okulov}
\email{alexey.okulov@gmail.com}
\homepage{http://okulov-official.narod.ru}
\affiliation{Russian Academy of Sciences, 
119991, Moscow, Russia}

\date{\ November 3, 2010}

\begin{abstract}
{ We consider the cold bosonic ensemble trapped by a helical 
interference pattern in the optical $loop{\:}$ scheme. 
This rotating helical potential is produced by 
the two slightly detuned counter propagating Laguerre-Gaussian 
laser beams with counter directed 
orbital angular momenta $\pm \ell\hbar $. The detuning $\delta \omega$ 
may occur due to rotational Doppler effect. 
The superfluid hydrodynamics is analysed for the large number 
of trapped atoms  in Thomas-Fermi approximation. For the highly elongated 
trap the Gross-Pitaevskii equation 
is solved in a slowly varying envelope approximation. 
The speed of axial translation and angular momenta of interacting 
atomic cloud are evaluated. In the $T \rightarrow 0$ 
limit the angular momentum of the helical cloud is expected to be 
zero while toroidal trapping geometry leads to $2\ell \hbar$ 
angular momentum per trapped atom.
}
\end{abstract}


\maketitle

\vspace{1cm}

\section{Introduction}
The hydrodynamics of the sufficiently 
cold ($T \sim 10^{-6} K$) bosonic ensemble 
trapped by 
optical  potential 
$V(\vec r,t)$ \cite{Cornell:2006,Cornell:2007,Ovchinnikov:2000} is 
described by the Gross-Pitaevskii equation (GPE) \cite{Pitaevskii:1999} 
for macroscopic  wavefunction ${\Psi (\vec r,t)}$ :
\begin{equation} 
\label{GPE5}
\ {{{i}} \hbar}{\:}{\frac {\partial {\Psi}}
{\partial t}} = 
 -{\frac {\hbar^2}{2 m}} \Delta \Psi + { V{(\vec r, t)}}{\:}
\Psi +{\frac{4{\pi}{\hbar}{\:}^2{\:}{a_s}}{m}}
|{\Psi}|{\:}^2 {\:} \Psi ,
\end{equation}
where $m$ is the mass of atom, $a_s$ is the two 
body s-wave scattering length. 
The negative $a_s$ reduces the energy of ensemble and 
causes the mutual attraction of atoms. This results 
in formation of bright solitons in $\bf {1D}$ and collapse 
in higher dimensions. On the contrary in repulsive BECs
atoms repel each other and dark solitons or vortices 
are formed. In the periodic 
potential gratings \cite{Arimondo:2002}: 
\begin{equation} 
\label{opt_lattice} 
V({\vec r},t)\sim  {I}{(z,r,t)}\sim
\exp{\:}[-r^2]{\:}\cos[{\:} \delta \omega t - (k_f+k_b) z ],
\end{equation}
where $r$ is a distance from propagation axis $z$,  
$ \delta \omega=c \cdot (k_f-k_b)=\omega_f-\omega_b$ is 
a frequency difference 
of the counter-propagating $z$-paraxial laser beams with the 
opposite wave vectors 
$|\vec k_{(f,b)}| \approx k_{(f,b)}$, 
the many-body nonlinearity 
leads to the nonlinear 
tunneling, self-trapping and other quantum interference 
phenomena \cite{Morsch:2006}. 
For accelerated $\bf {1D}$ optical gratings when 
$\delta \omega = \dot{\delta \omega} \cdot t$, i.e. in the  
non inertial reference frames \cite{Arimondo:2002} 
the interacting bosons demonstrate Bloch oscillations and 
Landau-Zener tunneling. In rotating $\bf {1D}$ lattices 
the repulsive ensemble may 
have stable ground states and vortex 
soliton states \cite {Kartashov:2007}. For the attractive 
ensemble in $\bf {2D}$ rotating 
lattices \cite {Sakaguchi:2007} the localized stable 
solitons exist in the certain range of angular velocities. Inside  
the parabolic well with trapping frequency $\omega_{\bot}$ and 
rotation frequency $\Omega$ the nonlinear 
localized matter waves appear when $\Omega \sim \omega_{\bot}$ at 
lowest Landau level \cite{Fetter:2009} which appears due to the $fictionary{\:}
magnetic {\:}field$ induced by trap rotation \cite {Sakaguchi:2008}.

The goal of the present work is to study bosonic ensemble in 
rotating reference frame by a virtue of rotating optical 
$dipole$ trap whose rotation is due to the 
frequency detuning $\delta \omega$. 
The proposed trap configuration is composed of the 
two counter propagating optical vortices. 
The shape of the 
interference pattern is defined by the mutual orientation 
of their orbital angular momenta $\pm \ell \hbar$ (OAM). 
When OAMs are 
co-directed the $\lambda/2$ 
spaced $toroidal$ traps are 
formed by Laguerre-Gaussian(LG) \cite{Okulov:2008} or Bessel 
vortices \cite{Sepulveda:2009}. 
This geometry had been used for the 
single atom trapping and 
detection \cite{Rempe:2007} and for 
the analysis of persistent condensate flows 
in LG beam waist \cite{Abraham:2002}. 
When OAMs are 
counter directed due to 
the phase-conjugation of the backward reflected 
LG beam the $\bf {1D}$ sinusoidal intensity 
grating is transformed into the 
truly $\bf {3D}$ $helicoidal$ (fig.\ref{fig.1},sec.II) 
grating $I(z,r,\theta,t)$ 
\cite{Okulov:2008,Sepulveda:2009,Bhattacharya:2007}. This 
grating experimentally observed for $\delta \omega =0$ 
\cite{Woerdemann:2009}:
\begin{equation} 
\label{twisted_lattice}
{I}{(z,r,\theta,t)}{\cong }r^{2|\ell|} \exp{\:}[-r^2]
\cos[{\:} \delta \omega t - (k_f+k_b) z +2\ell \theta],
\end{equation}
where $z,r,\theta$ are cylindrical coordinates, must rotate 
with angular frequency $\Omega= \dot \theta=\delta \omega /2 \ell$ when 
 $\delta \omega \ne 0$. 

Noteworthy the similar interferometric configuration with 
counter directed $spin$ optical angular momenta 
(circular polarizations) which is 
used for the sub-Doppler polarization gradient 
cooling \cite {Dalibard:1989}. The interference pattern 
forms the static potential 
gratings due to Zeeman shift between ground-state 
sublevels. This causes the 
Sizyphus deceleration of the atomic beam in optical molasses. 

We consider two possible regimes 
of the cold ensemble 
trapping when kinetic energy is small 
compared to the interaction and 
trapping energies. One Thomas-Fermi solution 
is obtained as a balance of the red-detuned optical 
attractive potential and 
self-defocusing due to positive $a_s$. This solution has 
atomic density $\rho_{_{h}}$ perfectly collocated 
with rotating optical helix $I(z,r,\theta,t)$.
In this case the rotating potential imposes 
rotation to superfluid. 
The other TFA solution has nonrotating 
density "funnel" profile $\rho_{_{fun}}$ relevant to thermodynamic 
limit when number of atoms $N \to \infty $. 

Next the analytical solution 
of GPE for nonzero 
kinetic energy $\Psi_{hel}$ is described. It is valid 
when optical trapping 
and interaction energies are comparable and subtract 
each other in GPE. 
The linear momentum $<P_z>$ and angular momentum 
$<L_z>$ of this helical atomic cloud are evaluated. 
The Landau 
criterion $|\vec V|> \epsilon(\vec p)/|\vec p{\:}|$ for 
the appearance of elementary excitations and 
superfluidity breakup 
\cite {Khalatnikov:2000} is discussed for helical geometry.

\section{Twisted wavetrains}
It is well known that interference of a two 
counter propagating waves with a different 
frequencies $\omega_f,\omega_b$ produces a 
running sinusoidal $roll$ intensity grating 
\cite{Morsch:2006,Zeldovich:1985}. 
For the equal wave amplitudes $|{\bf E}_f|$,$|{\bf E}_b|$ 
and the phase difference $\phi$ the distribution of 
the light intensity $I(z,r,\theta,t)$ has the 
following form: 
\begin{eqnarray}
\label{inter_patt2}
{I}{(z,r,\theta,t)} \sim
2 |{{\bf E}_{(f,b)}}|^2[1+ {\:}
\cos[  {\:} \delta \omega  t - (k_f+k_b) z+\phi {\:}]] 
&& \nonumber \\
\exp [ - 
{\frac {r^2}{{D_0}^2(1{+}z^2/z_R^2)}} ], {\:}{\:}{\:}
z_R=k_{(f,b)} {D_0}^2{\:}, {\:} 
{\:}{\:}{\:}{\:}{\:}
\end{eqnarray}
provided that a visibility of pattern is good 
enough $|{{\bf E}_{(f)}}|\cong|{{\bf E}_{(b)}}|$ 
\cite{Basov:1980}, where $z_R$ is Rayleigh range, $D_0$ is a beam waist radius,
the self-similar variable $\chi = (\omega_f-\omega_b) t - (k_f+k_b) z + \phi]$ is 
responsible for the translation of the interference pattern along $z$ axis with 
the $group$ velocity $V_z = (\omega_f-\omega_b)/(k_f+k_b)$. The 
transversal (in the plane $(r,\theta)$) confinement of the 
light amplitudes ${\bf E}_f,{\bf E_b}$ is typical 
for the zeroth-order Gaussian beams. 
The $roll$ interference pattern evolves into 
the sequence of the equidistantly spaced 
rotationally invariant in $\theta$ $ellipsoids$ centered at the propagation 
axis $z$ \cite{Okulov:2008}.
\begin{figure}
\resizebox{0.30\textwidth}{!}
{\includegraphics {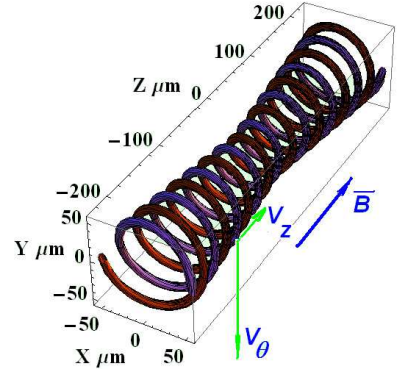}}
\caption{The isosurface of the optical intensity $I_{_{tw}}$ 
and the Thomas-Fermi density  ${\rho}_{h}$ of the cold atomic 
cloud in a helical 
optical dipole trap (\ref{inter_patt11},\ref{density_from_TFA})
( the scales are in 
$\mu m$, but longitudinal modulation of $\lambda/2$ is enlarged).
The spatial modulation 
is induced by the interference of 
counter-propagating LG beams with the 
opposite angular momenta. The pattern rotates with angular 
frequency $\delta \omega/2 \ell$ as a "solid-body". 
Magnetic field $\vec {\bf B}$ adjusts 
the scattering length $a_s$ to balance the 
attractive optical potential 
by many body defocusing.} 
\label{fig.1}
\end{figure}
For the higher-order propagation modes namely Laguerre-Gaussian beams (LG) with 
azimuthal quantum number $\ell$ and orbital 
angular momentum $\ell \hbar$ per 
photon \cite{Barnett:2002}: 
\begin{eqnarray}
\label{pump1}
{{\bf E}_{(f,b)}(z,r,\theta,t)} 
\sim {\frac {{{\bf E}_{(f,b)}} \exp [ i( -\omega_{(f,b)}t 
\pm k_{(f,b)} z) \pm i{\ell}\theta]} 
{ {(1{+}iz/z_R)}} } 
&& \nonumber \\
{\:}{({r}/{D_0})^{|\ell|}} 
\exp  {\:} [ {\:} - {\frac {r^2}{{D_0}^2(1{+}iz/z_R)}} ]{\:},
\end{eqnarray}
the interference pattern is different for LG reflected from conventional mirror and 
phase conjugating mirror. Backward reflection from conventional spherical mirror changes the 
topological charge of the LG \cite{Okulov:2008}, exactly in the same way 
as circular polarization of light changes from 
left to right and vice-versa in reflection \cite{Zeldovich:1985}.
The intensity $I_{_{tor}}{(z,r,\theta,t)}$ 
vanishes on the beam axis thus  
interference pattern transforms into the sequence 
of the equidistant rotationally invariant $toroids$ 
separated by $\lambda /2$ interval:
\begin{eqnarray}
\label{inter_patt4}
{I_{_{tor}}}{(z,r,\theta,t )} = A_{n}
{[1 + \cos[ \delta \omega \cdot t 
- (k_f+k_b) z]]}{({r}/{D_0})^{2|\ell|}}
&& \nonumber \\
\exp [ - {\frac {{2 r^2}}{{{D_0}^2}
(1+z^2/z_R^2)}} ], 
A_{n}={\epsilon_{_0} c}
{\frac {2 |{{\bf E}_{(f,b)}}|^2 {2^{(|\ell | +1)}}}
{\pi {\ell !}{D_0}^2}}
{\:}, 
\end{eqnarray}
where $\epsilon_{_0}$ is 
the dielectric permittivity of vacuum. 
The reflection from phase-conjugating 
mirror (PCM) 
does not change the topological charge of LG 
and the interference 
pattern is twisted \cite{Okulov:2008,Woerdemann:2009}: 
\begin{eqnarray}
\label{inter_patt11}
{I_{_{tw}}}{(z,r,\theta,t )} = 
{[ 1 + {\:}\cos[ {\:} \delta \omega \cdot t - (k_f+k_b) z + 
{2\ell}{\:}\theta ]]} \times 
&& \nonumber \\
A_{n} \cdot {({r}/{D_0})^{2|\ell|}}{\:}
\exp{\:}[-
{\frac {2 r^2}{{D_0}^2(1+z^2/z_R^2 )}}]
{\:}.{\:}{\:}{\:}{\:}{\:}{\:}{\:}
\end{eqnarray}
The intensity also vanishes at LG axis $z$ 
as $r^{2 |\ell|}$, while a self-similar argument:
\begin{equation} 
\label{self-similar}
\chi ={\:}[ {\:} (\omega_f-\omega_b) t - (k_f+k_b) z+2 {\ell}{\:}\theta {\:}],
\end{equation}
keeps 
the maxima of intensity at the $2\ell$ collocated helices 
separated from each other by $\lambda /2$ interval (fig. \ref{fig.1}).
The azimuthal term $2 \ell {\:} \theta$ appears due to phase 
conjugation ${{\bf E}_{{\:}b}} \sim {{\bf E}_{f}}^{*}$.
Thus we have the following strict correspondence between the $roll$ interference pattern 
(\ref{opt_lattice}) and the helical interference pattern (\ref{inter_patt11}): 
the frequency difference $\delta \omega=\omega_f - \omega_b$ is the cause of the translation 
of $rolls$ with $group$ velocity $V_z=(\omega_f - \omega_b)/ (k_f + k_b)$ of the wavetrain, 
produced by a sum of the two counter-propagating 
beams $({\bf E}_f+{\bf E}_b)$ \cite{Arimondo:2002}. The 
$\delta \omega$ is responsible also for the rotation 
of $helices$ with angular velocity $\dot \theta =\delta \omega /2 \ell$. The 
rotation is the cause of the $pitch$ of helical interference maxima along $z$-axis. 
Consequently there exists a perfect mechanical analogy 
between the $solid$ $body$ rotation of the helix 
described by formula (\ref{inter_patt11}) and an Archimedean screw. 
Namely the positive $\delta \omega$ corresponds to the counter-clockwise rotation and this 
provides the $pitch$ in positive $z$ direction for $right$ 
helices. On the other hand the negative $\delta \omega$ means clockwise rotation. In this case 
( $\delta \omega < 0$) the positive translation speed in $z$ direction takes place 
for the $left$-handed helices. 
Evidently the change of the topological charge $\ell$ changes the direction of 
helix translation $\vec V_z $ due to alternation of the helix hand to the opposite 
one for a given $\delta \omega$.

This mechanical analogy is useful for the analysis of the cold atoms motion 
in the helical trap. The velocity vector 
of the condensate fragment trapped and perfectly collocated with 
intensity maxima has two components 
$\vec V=\vec V_z + \vec V_{\theta}$(fig. \ref{fig.1}). The axial component is a $group$ velocity 
$|\vec{V}_z|=(\omega_f - \omega_b)/ (k_f + k_b)$ of the wavetrain,
while the azimuthal component 
$|\vec V_{\theta}|=(\omega_f - \omega_b)\cdot D_0$ is of kinematic nature. 

The electromagnetic orbital angular momentum inside LG 
beam waist volume $V \cong {\pi D_0^2 z_R}$ located near 
$z=0$ plane 
within Rayleigh range $|z|<z_R$ is the 
expectation value of 
the angular momentum operator 
$\hat L_z=-i{\hbar}{\:} [ {\:}{\vec r} \times 
{\nabla}] =-i{\hbar}{\:} {\frac {\partial }{\partial {\:} \theta}} {\:}$  
\cite{Sipe:1995}: 
\begin{eqnarray}
\label{OAM}
<L_z>_{(f,b)}=<{\Psi_{(f,b)}^{\ell}}|{\hat L_z}|{\Psi_{(f,b)}^{\ell}}> 
&& \nonumber \\
=2 \epsilon_0 {\int_V ({E^{+}_{(f,b)}})^{*}(-i{\hbar}{\:} [ {\:}{\vec r} \times 
{\nabla}]} {E^{+}_{(f,b)}}){\:}d^3 {\vec r} 
= 2 \epsilon_0 \times
&& \nonumber \\
{\int ({E^{+}_{(f,b)}})^{*} (-i{\hbar} {\frac {\partial }
{\partial {\:} \theta}}}{E^{+}_{(f,b)}}){\:}rdr{\cdot}d{\theta}dz 
\simeq {\pm \ell} {\hbar} {\frac {I_{(f,b)}V}{ \hbar \omega_{(f,b)}c}},
\end{eqnarray} 
where $I_{(f,b)}={\epsilon_{_0}}c |{{\bf{E}}_{(f,b)}}|^2$ is 
the light intensity, 
${\Psi_{(f,b)}^{\ell}} = \sqrt{2\epsilon_{_0}}{\:} {{E^{+}_{(f,b)}}{(z,r,\theta,t)}}$ 
are the macroscopic wavefunctions of a single 
photon inside a forward or backward beam (\ref{pump1}) 
with the winding number $\ell$, $E^+$ means the positive frequency 
components in $E(\vec r,t)$ spectrum. 
The square modulus $|\Psi_{(f,b)}^{\ell}|^2$ is a 
probability density of the photon detection which 
is proportional to the energy density of classical wave \cite{Sipe:1995}.
In this particular paraxial case the spin-orbit 
coupling \cite{Allen:1992} is small enough 
and the angular momentum of the photon is exactly 
decoupled to the 
spin and the orbital component: $\hat J = \hat S+ \hat L$. 
The linear momentum expectation 
values for the forward and 
backward LG are 
as follows:
\begin{eqnarray}
\label{momentum}
<P_z>_{(f,b)}=<{\Psi_{(f,b)}^{\ell}}|{\hat P}|{\Psi_{(f,b)}^{\ell}}>
=2 \epsilon_0 \times {\:}{\:}{\:}{\:}{\:} 
& \nonumber \\
{\int ({E^{+}_{(f,b)}})^{*}(-i{\hbar} {\frac {\partial }
{\partial z}}}{E^{+}_{(f,b)}}){\:}d^3 {\vec r} \simeq
{\:}{\pm \hbar}{\:} k_{(f,b)}
{\frac {I_{(f,b)}V}{\hbar \omega_{(f,b)}c}}. 
{\:}{\:}{\:}{\:}{\:}
\end{eqnarray} 
The ratio of the angular and linear momenta 
is $L_z/P_z \sim \ell{\:} c/{\omega_{(f,b)}}$ 
\cite{Allen:1992}.The angular and linear momenta of the composite 
wavetrains (\ref{inter_patt4}) and (\ref{inter_patt11}) (fig. \ref{fig.1}) are:
\begin{eqnarray}
\label{OAM_train}
<L_z>_{(f+b)}=<{\Psi_{f}^{\ell}}+{\Psi_{b}^{\pm \ell}}|{\hat L_z}|
{\Psi_{f}^{\ell}}+{\Psi_{b}^{\pm \ell}}> 
&& \nonumber \\
\simeq ({\ell} {\hbar} {\pm \ell} {\hbar} ) {\frac {I_{(f,b)}V}
{ \hbar \omega_{(f,b)}c}},
&& \nonumber \\
<P_z>_{(f+b)}=<{\Psi_{f}^{\ell}}+{\Psi_{b}^{\pm \ell}}|{\hat P}|
{\Psi_{f}^{\ell}}+{\Psi_{b}^{\pm \ell}}>
&& \nonumber \\
\simeq ({\hbar} k_f - {\hbar} k_b ) {\frac {I_{(f,b)}V}
{ \hbar \omega_{(f,b)}c}},
\end{eqnarray} 
where the upper sign in $\pm$ corresponds to reflection 
from conventional mirror while the bottom sign stands for the 
reflection from PC-mirror with alternation of the photon 
angular momentum.

The simplest conceivable configuration of linear PCM composed
of plane mirrors, wavefront curvature compensating lenses 
and beam splitters is described 
in \cite{Okulov:2010josa,Okulov_plasma:2010}. The goal of 
proposed setup is to counter-direct the splitted LG beams. The four 
reflections are the necessary minimum. The case is that the incidence 
angles much above $45$ degrees will distort both polarization 
and spatial structure of LG. In addition a 
small sliding of a beam along 
the reflecting surface \cite{Bliokh:2006} may occur 
when mirror is tilted with respect to 
the LG propagation axis. The evident physical restriction 
on this $loop$ setup is to keep the path difference 
$\Delta L$ of the counter directed LG smaller than coherence 
length of trapping 
laser field ($\Delta L<<c {\cdot}\tau_{coh}$) 
\cite{Okulov:2010josa}.The small frequency shift 
${\:} \delta \omega{\:} \approx 2\pi \cdot 10^{-(1-3)} rad/sec$ 
required to cause the helix rotation \cite{Okulov:2008} 
might be induced by a frequency ramp \cite{Arimondo:2002} or 
via rotational Doppler shift 
which appears due to rotation of the 
half-wavelength plate \cite {Dholakia:2002D} or 
Dove prism \cite {Padgett:1998}. 

\section{Cold ensemble density and velocity field in the helical trap} 
Consider a bosonic cloud prepared in an 
elongated trap \cite{Arimondo:2002} and suddenly 
released afterwards. The well elaborated 
experimental procedure is to 
impose a periodic optical potential to 
study the Bloch oscillations, macroscopic 
Landau-Zener tunneling and Josephson 
effects \cite{Morsch:2006,Pitaevskii:1999}. 
In our case the imposed optical potential is a $helical$ one: 
\begin{eqnarray}
\label{TRAPOP1}
\ { V_{opt}{(z,r,\theta,t)}}=- {\frac {Re[{\alpha(\omega)}]}
{2 \epsilon_{_0} c} {\:}
} I_{_{tw}}{(z,r,\theta,t)}, {\:}{\:}{\:}{\:}{\:}{\:}
&& \nonumber \\
\alpha(\omega) = 6 \pi \epsilon_{_0} {\:} c^3  
{\frac {\Gamma /{\:}{\omega_0}^2
 {\:} } 
{(\omega_0^2-\omega^2-{{i}}(\omega^3 / {\omega_0}^2)\Gamma)}},
\end{eqnarray}
where $\alpha(\omega)$ is the polarizability of atom, which is  
real, i.e. $\alpha(\omega) \approx Re[{\alpha}]$ at large 
detunings from resonance $\omega - \omega_0$, 
$\Gamma = e^2 {\omega_0}^2/6\pi {\epsilon_0 m_e c^3}$ 
is classical damping rate via radiative energy loss,
$m_e$ is electron mass 
\cite{Ovchinnikov:2000}. 
The GPE for wavefunction $\Psi$ of ensemble 
confined by $V_{opt}$ is:
\begin{equation}
\label{GPE5}
\ {{{i}} \hbar}{\:}{\frac {\partial {\Psi}(\vec{r},t )}
{\partial t}} = 
 -{\frac {\hbar^2}{2 m}} \Delta \Psi + { V_{opt}} {\:}
\Psi +{\frac{4{\pi}{\hbar}{\:}^2{\:}{a_s({\vec{B}})}}{m}}
|{\Psi}|{\:}^2 \Psi {\:},
\end{equation}
with $a_s({\vec{B}})=a_{bg}(1+{\Delta_B}/(|\vec{B}|-B_F))$ 
magnetic field dependent $s-wave$ scattering length, 
where $a_{bg}$ is background value 
of $a_s$, $B_F$ and $\Delta_B$ are the 
Feshbach magnetic induction and resonance width 
respectively. 

Consider the sufficiently large number of trapped 
atoms ($N \cong 10^{6-12}$). Then 
quantum pressure term 
following from the uncertainty principle is small 
(${\hbar^2}\Delta \Psi /2m \sim 0$) 
compared to the optical 
trapping and interaction terms \cite{Pitaevskii:1999}:  
\begin{equation}
\label{kin_int_ratio}
{\frac {E_{int}}{E_{kin}}} \cong {\frac {N \cdot a_{bg}}{a_{\bf {ho}} }}, 
{\:}{\:}{\:}{\:}
\end{equation}
where $a_{\bf{ho}}\cong {\sqrt{\hbar /m \omega_z}}$ 
is the "harmonic oscillator 
width" \cite{Pitaevskii:1999}. 
The ratio $E_{int}/E_{kin}$ (\ref{kin_int_ratio}) 
is much more than unity 
for the most of the near infrared lasers $\lambda=0.8-1.5 \mu m$, 
under the 
standard focusing requirement $D_0 \sim 10 - 100 \mu m$ and 
when $a_s(\vec B)$ is tuned in the range
$ \cong 1 - 100 nm$ via Feshbach 
resonance. For the mid-infrared trapping at 
$CO_2$ lasing wavelength 
$\lambda=10.6 \mu m$
the kinetic energy term $E_{kin}=2 \hbar^2 /{m \lambda^2}$ 
is about $100$ times smaller. The optical dipole 
trapping energy $E_{dipole}$ is:
\begin{eqnarray}
\label{dipole_energy}
{E_{dipole}}\cong N \cdot V_{opt}\cong -{\frac {\vec p \cdot\ \vec E}{2}}
 \cong {\frac {- \alpha \vec E^2 N}{2}} 
 && \nonumber \\
\cong {\frac {- \alpha I_{tw} N}{2 \epsilon_0 c}}=
{\frac {- e^2 I_{tw} N}{m_e \Delta \omega^2 2 \epsilon_0 c}}, {\:}\Delta \omega = \omega_{f,b} - \omega_{0}
{\:}{\:}{\:}{\:}
\end{eqnarray}
where $I_{tw}$ is the optical 
intensity of trapping beams \cite{Ovchinnikov:2000}. 
Hence one might expect 
that a Thomas-Fermi approximation (TFA) is adequate 
in our case when interaction is repulsive 
$a_s>0$ \cite{Pitaevskii:1999}.

When inhomogeneous ensemble is considered in the 
thermodynamic limit ($N \to\infty $) the local 
density approximation for chemical potential 
is used $\mu(r)=\mu_{_{local}}[\rho(\vec r)]+V_{opt}(\vec r)$ 
\cite{Pitaevskii:1999} .  
Let us consider the following 
TFA wavefunction in the vicinity of the 
LG-beam waist (i.e. within Rayleigh range $|z|<z_R$):
\begin{equation}
\label{TFA_wavefunction}
\Psi{_{fun}} =
\Phi (r) \exp{\:}[- {\frac {i \mu(r) t} {\hbar}} + 
i{\Phi(r)}^2 
\sin(\delta \omega t - 2k_z z+2 \ell \theta)],
\end{equation}
where the local $r$-dependent 
chemical potential $\mu (r)$ is:
\begin{eqnarray}
\label{TFA_wavefunction_parameters1}
{\mu (r)}= 4 \pi \hbar^2 a_s \Phi(r)^2 /m ;{\:}
{\:}I_{tw}=2 \epsilon_{_0} c |{\bf E}_{(f,b)}|^2; {\:}{\:} 
&& \nonumber \\
{\Phi(r)}^2 = 
\exp(-2r^2/D_0^2) \cdot {(r/D_0)^{2|\ell|}} 
\cdot{\frac {\alpha(\omega)  I_{tw}}{2 \epsilon_{_0} c 
\cdot \hbar \cdot {\delta \omega}}}.
\end{eqnarray}
The wavefunction (\ref{TFA_wavefunction}) is 
normalizable and fits the GPE 
by substitution. This TFA density of atoms 
$\rho{_{_{fun}}}(z,r,\theta,t)=
|\Psi{_{fun}}|^2\sim \exp(-2({r/D_0})^2)/({r/D_0})^{2|\ell|}$ 
is $(z,t)-$ independent "funnel" collocated with the optical 
helix $I_{_{tw}}{(z,r,\theta,t)}$. 
The phase modulation of $\Psi_{fun}$ has a maximum near the density 
maximum $\rho_{_{fun}}(z,r,\theta,t)$, sinusoidal  
dependence on azimuthal angle $\theta$ and decreases down 
to zero on LG axis and outside 
the LG waist. Noteworthy $\Psi_{_{fun}}$ is a multiply 
valued function of $\theta$. Hence this solution is of 
restricted interest. It may be used for evaluations of the 
thermodynamical parameters of the cold ensemble with 
$|\Psi_{fun}|^2$ density \cite{Pitaevskii:1999}. 

The other solution 
for dilute Bose gas (e.g. $N \sim 10^6$) 
is obtained in the TFA for $r$-independent chemical potential 
$\mu(\vec r)=const$. 
The mean field wavefunction here $\Psi_h$ is a sum 
of the two phase-conjugated vortices with the opposite 
angular momenta $\pm \hbar \ell$: 
\begin{eqnarray}
\label{TFA_wavefunction_2}
\Psi_{_h} (z,r,\theta, t) =
\Psi_{\ell} (z,r,\theta, t)+ \Psi_{-\ell} (z,r,\theta, t) 
\cong {\:}{\:}{\:}{\:}{\:}{\:}{\:}{\:}{\:}{\:}{\:}{\:}{\:}
&& \nonumber \\
\Psi_{\pm \ell}(z=0){\cdot}{\frac{({r}/{D_0})^{|\ell|}}{1+z^2/z_R^2}}{\:}
\exp  {\:} \bigg{[} {\:} - 
{\frac {r^2}{{D_0}^2(1{+}iz/z_R)}} \bigg{]}\times 
{\:}{\:}{\:}{\:}{\:}{\:}{\:}{\:}{\:}
&& \nonumber \\
\bigg{\lbrace} 
{\frac{\exp[ {-{\frac {i \mu_{_f} t}{\hbar}} + i k_f z+i\ell \theta} ]}
{{(1+iz/z_R)}}}+
 {\frac {\exp[ {-{\frac {i \mu_{_b} t}{\hbar}} - ik_b z-i\ell \theta} ]}
{{(1+iz/z_R)}}}
 \bigg{\rbrace} ,{\:}{\:}
\end{eqnarray}
where the difference of the partial chemical 
potentials $(\mu_{_f}-\mu_{_b})$,
associated with the each of "counter-propagating" 
wavefunctions $\Psi_{\ell},\Psi_{-\ell}$
is adjusted to the frequency difference of counter-propagating 
optical fields 
$(\mu_{_f}-\mu_{_b})/{\hbar}=\delta {\:} \omega 
=\omega_f - \omega_b$.
The substitution of this TFA wavefunction into GPE 
gives the following link 
for parameters: 
\begin{eqnarray}
\label{density_from_TFA}
\mu - {\frac {\alpha(\omega){{I_{_{tw}}}(0)}{\cdot}
[1+\cos(\delta \omega t + 2kz \pm 2\ell \theta) ]}
{{2 \epsilon_{_0} c}\cdot 
g {\:}(1+z^2/z_R^2)}}{(r/D_0)^{2|\ell|}}\times 
&& \nonumber \\
\exp  {\:} \bigg{[} {\:} - 
{\frac {2{\:} r^2}{{D_0}^2(1+z^2/z_R^2)}}\bigg{]}=
\rho_{h}(z,r,\theta,t),{\:}{\:}{\:}
{\:}{\:}{\:}{\:}{\:}
\end{eqnarray}
where $k=k_f \cong k_b$,
$\mu=\mu_{_f} \approx \mu_{_b}$ is a constant 
($z,r,\theta,t$-independent)
 value of chemical potential, 
${g}={{4 \pi \hbar^2 a_s({\vec{B}})}/{m}}$ is 
the interaction parameter. 
This is the quasiclassical restriction 
imposed on a homogeneity 
of chemical potential of the system 
in an external field $V_{opt}$ \cite{Pitaevskii:1999}. 
In accordance to this solution 
the density of the cold atomic 
ensemble $\rho_{h}(z,r,\theta,t)$ is perfectly correlated 
with the rotating optical helix 
potential $I_{_{tw}}(z,r,\theta,t)$ as depicted at fig.\ref{fig.1}.
The density $\rho_{h}$ rotates as a "solid body". 
The speed of the axial translation 
is $V_z= \lambda \cdot \delta \omega/4\pi$. 

Apart from TF approximation the truly exact solution 
with nonzero kinetic energy exists 
$\Psi_{ex}=\Psi_f+\Psi_b$. It is also a superposition of the two 
counter propagating $paraxial$ matter waves $\Psi_f$ and $\Psi_b$. 
Let us reduce GPE (\ref{GPE5}) to the $paraxial$ form 
relevant to highly elongated geometry, as a $helical$ one in our case:
\begin{eqnarray}
\label{GPE_paraxial}
{i \hbar}{\frac {\partial {\Psi_{(f,b)}}}
{\partial t}} = -{\frac {\hbar^2}{2 m}} 
\Delta_{\bot} {\Psi_{(f,b)}} 
-i {\frac {{2k_{(f,b)}} \hbar^2}{2 m}} {\:} 
{\frac {{\:} \partial \Psi_{(f,b)}}{\partial z}} 
&& \nonumber \\
+ { V_{opt}}{\:}{\Psi_{(f,b)}} +{\frac {\hbar^2}{2 m}}{\:} {{k^2}_{(f,b)}} {\Psi_{(f,b)}} 
+{g}
|{{\Psi_{f}} +{\Psi_{b}} }|{\:}^2 {\Psi_{(f,b)}},
\end{eqnarray}
The perfect mutual cancellation of trapping 
and interaction terms:
\begin{equation}
\label{GPE_paraxial1}
-{ V_{opt}{(z,r,\theta,t)}}{\:}{\Psi_{(f,b)}} 
={g}
|{{\Psi_{f}} +{\Psi_{b}} }|{\:}^2 {\Psi_{(f,b)}} ,{\:}
\end{equation}
occurs when nonlinear defocusing 
due to positive scattering length $a_s$ is compensated by 
attraction to intensity maxima caused by red detuning.
Taking again the exact helical solution of (\ref{GPE_paraxial})
as a superposition of the two counter-propagating 
vortices:
\begin{eqnarray}
\label{wavefunction_exact}
\Psi_{_{ex}} (z,r,\theta, t) =
 \Psi_{f} (z,r,\theta, t)+ \Psi_{b} (z,r,\theta, t) 
\cong {\:}{\:}{\:}{\:}{\:}{\:}{\:}
&& \nonumber \\
{\tilde \Psi_{f}}{\cdot}{\exp[ {- {\frac {i \mu_{_f} t}{\hbar}} + i k_f z} ]}
+
{\tilde \Psi_{b}}{\cdot}{\exp[ {- {\frac {i \mu_{_b} t}{\hbar}} - i k_b z} ]},
\end{eqnarray}
and under the natural assumptions: 
\begin{equation}
\label{paraxial assump}
k_{(f,b)}{\:} 
{\partial \tilde \Psi_{(f,b)}}/{ \partial z}>> 
{{\:} \partial^2 }{\tilde \Psi_{(f,b)}}/{ \partial z}^2 ,
\end{equation}
two following equations for counter propagating and counter rotating 
matter waves $\tilde \Psi_{f}$ and $\tilde \Psi_{b}$ are valid: 
\begin{equation}
\label{paraxial shrod}
i 2k_{(f,b)}
{\frac {\partial \tilde \Psi_{(f,b)}}{ \partial z}}+
{\Delta_{\bot} \tilde \Psi_{(f,b)}}
-(k^2_{(f,b)}+{\frac {2 m \mu_{(f,b)}}{\hbar^2}} )  \Psi_{(f,b)}=0, 
{\:}   
\end{equation}
which have the vortex solutions with charge $\ell$ 
for the ${\tilde \Psi_{f,b}}(z,r,\theta)$ 
with initial condition at 
$z=0$ equals ${\tilde \Psi_0}$:
\begin{equation}
\label{wavefunc_parax}
{\tilde \Psi_{(f,b)}} \sim
{\frac{{\tilde \Psi_0}{\cdot}{({r}/{D_0})^{|\ell|}} {\cdot}
\exp{\:} \bigg{[} {\:}-
{\frac {r^2}{{D_0}^2(1{+}iz/{z_R}))}
\pm i \ell \theta} \bigg{]}}
{{{(1+iz/ { z_R})}}}}{\:}.
\end{equation}
The issues 
of dynamical stability of (\ref{wavefunction_exact}) 
with respect to small perturbations and thermodynamical 
stability from the point of view of least energy arguments deserve a further 
careful analysis and will be published elsewhere. 
For example transformation to the reference frame rotating 
with angular velocity $\Omega=\delta \omega /2 \ell$ synchronously 
with trapping helix leads to equation 
\cite {Pitaevskii:1999,Fetter:2009,Sakaguchi:2008}:
\begin{eqnarray}
\label{GPE_rot_frame}
{i \hbar}{\frac {\partial {\Psi}}
{\partial t}} = -{\frac {\hbar^2}{2 m}} 
\Delta {\Psi} +{ \tilde V_{opt}{(z,r,\theta)}}{\:}{\Psi} 
+{g}
|{\Psi}|{\:}^2 {\Psi}-\Omega \hat L_z {\Psi},
&& \nonumber \\
\tilde V_{opt}\sim r^{2|\ell|} 
\exp({\frac {-2r^2}{D_0^2(1+{ {z^2}/{k^2 D_0^4}})}})
[1+\cos(2kz+2\ell\theta)],{\:}{\:}
\end{eqnarray}
where stationary solutions for the diluted ($\mu=const$)
ensemble $\Psi=\Phi(z,r,\theta)\exp(-i\mu t/ \hbar)$ 
are given by:  
\begin{equation}
\label{GPE_static_in_rot_frame}
\mu {\Phi} = -{\frac {\hbar^2}{2 m}} 
\Delta {\Phi} +{\tilde  V_{opt}{(z,r,\theta)}}{\:}{\Phi} 
+{g}|{\Phi}|{\:}^2 {\Phi}+
\Omega {i \hbar}{\frac {\partial {\Phi}}{\partial \theta}}.{\:}
\end{equation}

\section{Macroscopic observables} 
The helical solutions 
composed of the counter propagating free space 
LG wavefunctions apparently fit the 
continuity equation and 
have realistic field of velocities. In addition to 
the ensemble density $\rho(\vec r,t)$ obtained above 
in TFA the structure of velocity field $\vec V(\vec r,t)$ 
is a consequence of the complex geometry of the helical 
wavetrain which requires the perfect adjustment of the 
phase fronts achieved by phase-conjugation of colliding vortices. 
The one possible application of the 
(\ref{TFA_wavefunction_2},
\ref{wavefunction_exact}) 
might be in using them as variational anzatz for 
emulation of GPE \cite{Malomed:2002}. 
Nevertheless the explicit form of solutions $\Psi_{h}$ 
(\ref{TFA_wavefunction_2},\ref{wavefunction_exact}) 
offers a possibility to evaluate 
the macroscopic observables of the 
trapped ensemble. Helical wavetrain has 
nonzero momentum $P_z$ (\ref {momentum}): 
\begin{eqnarray}
\label{momentum2}
<P_z>_{h}=<{\Psi_{h}}|{ -i{\hbar}
{\frac {\partial }
{\partial z}}}|{\Psi_{h}}> =N {\hbar} (k_f - k_b)
\Leftarrow  \hbar \int_V dV \cdot
&& \nonumber \\
\exp(-r^2)r^{2|\ell|}[ k_f - k_b +
k_f \exp(i \chi) - k_b \exp(-i \chi) ],{\:}{\:}{\:}
{\:}{\:}{\:}{\:}{\:}{\:}
\end{eqnarray}
and easily calculated angular momentum $L_z$ (\ref {OAM}): 
\begin{eqnarray}
\label{OAM1}
<L_z>_{h}=<{\Psi_{h}}|-i{\hbar}{\frac {\partial }
{\partial {\:} \theta}}|{\Psi_{h}}>=
N{\ell \hbar}(1\mp1) \Leftarrow{\:}{\:}{\:}{\:} 
&& \nonumber \\
\hbar \int_V {dV \exp(-r^2)r^{2|\ell|}[1 \mp 1+
\exp(i \chi) \mp \exp(-i \chi)]},{\:}{\:}{\:}{\:}{\:}
{\:}{\:}{\:}
\end{eqnarray} 
due to the apparent identity 
$\int^{2\pi}_0 \sin (\chi)d {\theta}=0$. The upper $\mp$ 
sign in (\ref {OAM1}) corresponds to counter 
directed angular momenta and 
helical interference pattern (\ref{inter_patt11}), 
while bottom $\mp$ 
sign corresponds to the toroidal optical interference 
pattern (\ref{inter_patt4}). 

The same expectation 
values $<P_z>_{ex}$=$N {\hbar} (k_f - k_b)$ 
and $<L_z>_{ex}= N \ell \hbar (1 \mp 1)$
has exact 
wavefunction $\Psi_{ex}$(\ref{wavefunction_exact}). 
Quantum mechanically this happens because the 
wavefunction 
in both cases is 
a superposition of the two partial matter 
waves $\Psi_f$ and $\Psi_b$ in 
(\ref{wavefunction_exact}) (or 
$\Psi_{\ell}$ and 
$\Psi_{-\ell}$ in (\ref{TFA_wavefunction_2})) having opposite 
and quantized (i.e. equal to $\pm \ell \hbar$) 
mutually subtracted 
angular momenta. This means also that helical optical wavetrain 
(\ref{inter_patt11}) contains 
OAM of exactly $0 \times \ell \hbar$ per photon, while 
toroidal wavetrain (\ref{inter_patt4}) 
contains $2 \ell \hbar$ per photon (\ref{OAM_train}) as shown in 
sec.II.

This feature looks seemingly counter intuitively from
the point of view of classical hydrodynamics, but the similar 
results on vanishing of the moment of inertia for 
purely superfluid ensemble were 
summarized in \cite {Pitaevskii:1999}. 
Namely the density 
of atomic ensemble 
$\rho_{h}$  =  $|\Psi_{h,ex}|^2 $ rotates as a $solid$ 
$body$ and one might expect that $\rho_{h}$ to 
have classically the 
angular momentum 
$L_{class}=I_{zz} \cdot \delta \omega /2 \ell$, 
where $I_{zz}$ is the moment of inertia 
of the helical $wire$ located in LG beam waist 
with the density 
profile $\rho_{h}$ \cite {Landau:1976}: 
\begin{eqnarray}
\label{inertia_tensor}
I_{zz}= \int Nm{|\Psi_{h,ex}}|^2 r^2 dV = 
Nm \int {|\Psi_{h,ex}}|^2 r^3 dr d{\theta} dz \simeq
&& \nonumber \\
\sim Nm \int (1+\cos(\delta \omega t+(k_f+k_b)z+2{\ell}{\theta}))
 \cdot d{\theta} dz \cdot
{\:} {\:} {\:} {\:} {\:} {\:} {\:} {\:} {\:} 
&& \nonumber \\
{\exp(-r^2/{D_0}^2)} r^{2+2|\ell|} \cdot rdr\sim Nm {D_0}^2 Z_r.{\:} {\:} {\:} 
{\:}{\:} {\:} 
\end{eqnarray}

Nevertheless due to the $quantization$ of the angular momentum in free space,  
the oppositely directed angular momenta cancel each other completely, 
because they have integer opposite values of $\pm \ell \hbar$ 
\cite {Phillips:2006}. In $T \rightarrow 0$ 
limit \cite{Pitaevskii:1999} the 
net angular momentum of the helical 
wavetrains (\ref{TFA_wavefunction_2},
\ref{wavefunction_exact})  
is zero because the superfluid component remains only. 
On the contrary, the linear momentum is $not$ quantized in free space 
and this leads to nonzero net linear momentum $P_z$ 
of the ensembles (\ref{TFA_wavefunction_2},\ref{wavefunction_exact}), regardless to the mutual 
orientation of their OAM's. The net 
momentum $P_z$ is small because of the smallness of the $group$ velocity 
of the helical wavetrain $V_{z}=\delta \omega / (k_f+k_b)$. For 
example when frequency splitting $\delta \omega$ is induced by 
rotational Doppler effect \cite{Okulov:2010josa,Dholakia:2002D}, 
the speed of the axial translation of the helical density profiles 
(\ref{TFA_wavefunction_2},\ref{wavefunction_exact}) $V_{z} \sim $ is 
several $\mu m$ per second (several 
rotations of helix per second). 

The else interesting physical consequences relevant to 
experiments with trapped quantum gas traps may 
be formulated  from the 
point of view of the Landau 
criterion $|\vec V|> \epsilon(\vec p)/|\vec p{\:}|$ for 
the appearance of elementary excitations (rotons) and superfluidity 
destruction, where $\epsilon(\vec p)$ is the energy - momentum 
dispersion relation. Following 
to \cite{Khalatnikov:2000,Feynman:1972} consider 
the flow of quantum gas in a narrow helical channel with 
velocity $\vec V=\vec V_z + \vec V_{\theta}$. 
In the rest frame the momentum of excitation $\vec p{\:}$ 
must be opposite to the velocity of superfluid $\vec V$, 
because of the least energy constraint imposed upon 
excitation $\epsilon(\vec p)+ \vec p \cdot \vec V<0$. 
Thus $\epsilon(\vec p)- |\vec p| \cdot |\vec V|<0$ and 
$|\vec V|> \epsilon(\vec p)/|\vec p{\:}|$. Because in our 
case the only significant component of $\vec V$ is 
$V_{\theta}=\delta \omega D_0 /2 \ell $ the excitations 
with momentum $\vec p$ appear when:
\begin{equation}
\label{super}
\delta \omega_{crit} D_0 > 2 \ell \cdot \epsilon(\vec p)/ |\vec p{\:}|.
{\:}{\:}{\:}{\:}{\:}
\end{equation} 
The experimentally controllable 
detuning $\delta \omega$ of counter 
propagating waves $\omega_f$ and $\omega_b$ by rotational 
Doppler effect \cite {Okulov:2010josa} which leads to the  
change the angular velocity of helix rotation 
makes possible to 
determine the critical velocity of superfluid, defined 
by contact point 
of roton minimum of $\epsilon (\vec p)$ 
with the line $|\vec p| V_{\theta}$.   
The turbulent excitations (rotons) are 
assumed to appear due to ejection of superfluid across 
the trapping potential barrier owing to centrifugal force,
rather than because of the roughness or the channel end 
\cite {Khalatnikov:2000,Feynman:1972}.
\section{Conclusion}
	The flow of the degenerate quantum gas in helical 
trap had been studied 
analytically in the framework of the Gross-Pitaevskii 
equation. The necessary conditions 
were formulated for the appearance of the 
helical Bose-Einstein condensate flows due to the 
Thomas-Fermi balance of the self-defocusing 
of condensate with positive scattering length $a_s$ 
and "red" detuned optical dipole potential. 
The minimal achievable ensemble temperature 
might be approximately evaluated as a recoil one 
$ T_{recoil}=4 \cdot \hbar^2/(2 m \lambda^2 \cdot k_B)$ 
\cite {Dalibard:1989}. The possible 
experimental implementation of helical trapping 
is a sudden $switching$ $on$ of the 
helical potential 
after the condensate release from elongated optical trap in 
a way similar to switching of accelerated grating 
in Ref. \cite{Arimondo:2002}. 

The peculiarities of the cooling mechanisms 
in this helical configuration were not 
considered in the current work. But 
the helical interference pattern (see fig.\ref{fig.1}) 
geometry might reveal the new features of 
the well elaborated mechanisms as a 
the Doppler cooling \cite {Letokhov:1968}, 
polarization gradient cooling \cite {Dalibard:1989} or 
velocity selective population trapping \cite {Aspect:1988}.
The newly found loop and helical features of the 
optical speckle patterns \cite{Padgett:2009,Okulov:2009} 
are also a promising trapping opportunities which 
may enlighten the features of the Anderson localization of cold atoms 
in $\bf 1D$ and $\bf 3D$ speckle patterns 
\cite {Inguscio:2008}. 

Noteworthy 
the similar helical geometry of the colliding LG optical vortices 
of picosecond duration 
with opposite angular momenta proposed recently for the 
plasma currents excitation via ponderomotive force 
\cite{Okulov_plasma:2010}. 
As the plasma vortices are the sources of the 
axial magnetic fields, 
the superfluid motion in helical trapping environment (\ref{GPE5})
is to be associated 
with a so-called $artificial$ 
magnetic fields \cite{Fetter:2009,Sakaguchi:2008}. 

The partial support of the Russian Fund for Basic Research 
through grant 08-02-01229 is acknowledged. 

\end{document}